\documentstyle[twocolumn,aps]{revtex}
\begin{document}

\title{
The Same Superconducting Criticality for Underdoped and Overdoped  La$_{2-x}$Sr$_x$CuO$_4$ Single Crystals
\vskip 0.25truecm
\normalsize
\noindent
H. H. Wen$^1$, X. H. Chen$^2$, W. L. Yang$^1$, and Z. X. Zhao$^1$
\vskip 0.1truecm
{\it
$^1$ National Laboratory for Superconductivity,
Institute of Physics and Center for Condensed Matter Physics,
Chinese Academy of Science, P.O. Box 603, Beijing 100080, P. R. China\\
$^2$ Physics Department, University of Sciences and Technology of China, Hefei 230026, P.R.China
}
\vskip 0.2 truecm
\small 
\begin{flushleft}
By measuring the superconducting diamagnetic moments for an underdoped 
and an overdoped La$_{2-x}$Sr$_x$CuO$_4$ single crystal with equal 
qualities and roughly equal transition temperatures, it is found that 
the underdoped sample has only one transition which corresponds to 
H$_{c2}$, but the overdoped sample has two transitions with the higher 
one at H$_{c2}$. Further investigation reveals the same upper-critical 
field H$_{c2}$ for both samples although the overall charge densities 
are very different, indicating the possibility of a very direct and 
detailed equivalence of the superconducting condensation process in 
the two doping limits. The second transition for the overdoped sample 
can be understood as the bulk coupling between the superconducting 
clusters produced by macroscopic phase separation.
\end{flushleft}
\small
\begin{center}
PACS numbers : 74.25. Bt, 74.20.Mn, 74.40.+k
\end{center}
}

\maketitle



The mechanism of high temperature superconductors ( HTS ), one of the 
challenging issues, has stimulated enormous effort in recent years. 
Connected with it is a widely accepted electronic phase diagram which 
contains three major phases: hole underdoped, optimally doped and overdoped. 
The contrasting properties in the normal state between an underdoped and 
an overdoped sample tempt to ascribe the superconductiviy to different 
condensation processes and thus different criticalities. One example is 
the recently proposed model of thinking the superconducting transition 
of HTS as a Bose-Einstein condensation in the underdoped region and the 
BCS-like origin in the overdoped region [1,2]. Therefore it remains unclear 
wether the HTS has the same condensation process when going through the u
nderdoped region to the overdoped region. Another problem arose in an 
overdoped HTS is that the transition temperature T$_c$ drops with increasing 
the number of charge carriers ( here the doped holes ), in sharp contrast 
with what appears in the underdoped region. The crossover from the non-Fermi 
liquid in the underdoped region to the Fermi liquid behavior in the overdoped 
region with increasing doping level clearly shows that most of the doped holes 
join the conduction in the normal state. Recent data from the measurement on 
the penetration depth $\lambda$ [3,4], show that, however, the superfluid 
density $\rho$$_s$ behaves just like the transition temperature T$_c$, i.e., 
decreases with the doped hole number.   The consequence is that in the overdoped 
region, the more charge carriers are doped, the less superfluid density $\rho$$_s$ 
will be.  Therefore, the doped holes in the overdoped region seem to be separated 
into two parts, only part of them condense into a lower energy state leading to 
the superconductivity. In our previous paper [5] it was shown that the macroscopic 
phase separation may have occurred in overdoped 
Bi$_{2}$Sr$_{2-x}$La$_{x}$CuO$_{6+y}$ single crystals although we were not sure whether this phase separation is induced by the inhomogenneity of excess oxygen or by 
the electronic instability. In this Letter, we present a comparative investigation 
on an underdoped and an overdoped single crystal. A similar two-step transition 
has been observed only in the overdoped sample leading to an intuitive 
inference that the macroscopic phase transition may have occurred in 
this sample due to the electronic instability rather than the chemical 
inhomogeneity since in this system the incorporation of excess oxygen 
is very difficult especially for the overdoped sample. 

Single crystals measured for this work were prepared by the traveling solvent 
floating-zone technique [6]. A series of single crystals have been investigated 
for this study. For the sake of simplicity, in this Letter we present the 
measurement only for two typical single crystals, one underdoped and another 
one overdoped with almost the same transition temperatures and equal 
qualities. Shown in Fig.1 are the superconducting transitions of these 
two typical samples with dimensions of about 
$2 mm ( length ) \times 1 mm ( width ) \times 0.3 mm ( thickness )$ 
measured by a superconducting quantum interference device ( SQUID, 
Quantum Design, MPMS 5.5 ). Resistive measurements on these samples 
show very narrow transition widths ( $<$ 1K )  indicating a high 
quality of the samples. The transition temperatures of the 
overdoped ( x = 0.24 ) and the underdoped ( x = 0.092 ) samples 
are 25 K and 26 K, respectively, which fall exactly onto the general 
parabolic curve of T$_c$ versus doping level with optimal doping at 
about 0.16 and T$_c$ = 38.5 K as found by many others [6, 7]. 
The similar qualities and transition temperatures between the 
underdoped and the overdoped samples provide us an effective 
way to do the comparative investigation. 

Distinct diamagnetic behaviors have been found and shown in Fig.2 for 
the underdoped and the overdoped samples when a relatively strong 
external magnetic field is applied. It is clear that for the underdoped 
sample, there is only one transition temperature marked here as T$_{c1}$. 
The slight diamagnetic moment appeared above T$_{c1}$ is due to the 
fluctuation effect. For the overdoped sample, however, there are two 
transitions, one appears at almost the same temperature as the underdoped 
sample, i.e., T$_{c1}$, while another sharp transition occurs at T$_{c2}$. 
The irreversibility for flux motion appears immediately after T$_{c1}$ 
for the underdoped sample and after T$_{c2}$ for the overdoped sample. 
It is further found that the first transition temperature T$_{c1}$ varies 
with the external magnetic field very slowly, while the second one 
T$_{c2}$ shifts dramatically. The behavior of two transitions on one 
single M(T) curve was previously found in 
Bi$_{2}$Sr$_{2-x}$La$_{x}$CuO$_{6+y}$ single crystals [5] in which the 
excess oxygen may be inhomogeneous and thus the first transition was 
attributed to the appearance of superconductivity on some individual 
clusters ( with less oxygen and / or holes ) and the second transition 
is due to the Josephson coupling or proximity effect between these 
clusters. The two transitions in our present overdoped sample can 
get the same explanation but clearly the superconducting clusters 
here are not formed by the inhomogeneity of excess oxygen, rather 
by the electronic phase separation effect on the holes. 

Although the underdoped and the overdoped samples investigated here have 
almost the same superconducting transition temperatures at zero field, 
it gives however no reason to believe that the two samples have the same 
criticality at a high magnetic field since they have very different overall 
hole densities. The criticality, such as the upper critical field B$_{c2}$(T), 
contains important information about the superconducting condensation 
and probably is also related to the pairing mechanism of Cooper pairs, 
therefore it is interesting to investigate the criticality of the underdoped 
and the overdoped samples. For this purpose, we determined the upper 
critical field B$_{c2}$(T) for both samples. For the underdoped sample, 
this is quite easy since there is only one sharp transition. For the 
overdoped sample, the reversible region is wide and the transition near 
T$_{c1}$ is rounded, therefore one should use the critical fluctuation 
theory [8] to derive the information of B$_{c2}$(T). In Fig.3, the 
temperature dependence of the reversible magnetic moments measured 
under 6 magnetic fields ( 0.2 T to 5 T ) for the overdoped sample are 
shown. There is a common crossing point at ( T$^*$, M$^*$ ) on these 
curves suggesting strongly an underlying scaling behavior. According 
to the fluctuation theory of Ullah and Dorsey [8], a general scaling 
law for high temperature superconductors ( HTS ) reads
\begin{equation}
{M \over (B_{ex}T)^\alpha} \propto G\left[{T-T_c(B_{ex}) \over (B_{ex}T)^\alpha}\right]
\end{equation}
where G(x) is an unknown scaling function, $\alpha$ = 2/3 for 3D and 1/2 for 2D, 
M is the magnetic moment, B$_{ex}$ is the external field. All the information 
about the upper critical field is included in the relation T$_c$(B$_{ex}$) or 
vice versa B$_{c2}$(T), namely a correct choice for the relation T$_c$(B$_{ex}$) 
will collapse all the M(T) curves onto one master line. Above scaling law has 
been well checked for various HTSs [9] delivering a high slope of B$_{c2}$(T) 
near T$_c$. As shown by the inset to Fig.3, by assuming 
B$_{c2}$(T) = ( T-T$_c$ )$\times$(dB$_{c2}$ / dT),   a good scaling can be 
obtained by taking  $\alpha$ = 2/3 ( 3D ) and dB$_{c2}$ / dT = - 0.7 $\pm$ 0.3 T / K. 
The upper critical field B$_{c2}$(T) for the overdoped sample determined by doing 
above scaling and that for the underdoped sample determined directly from the 
sharp transition at T$_{c1}$ are plotted together in Fig.4. It is remarkable 
that both curves are very close to each other. {\it This is our central result 
which indicates the same superconducting criticality for the underdoped and the 
overdoped samples.} It is important to note that for the underdoped sample, the 
correct way to determine B$_{c2}$ is also to do the critical scaling. Since in 
our present sample, the fluctuation region is too small to do that, therefore 
we determined the B$_{c2}$ directly from the sharp transition at T$_{c1}$. For 
underdoped YBa$_2$Cu$_3$O$_{7-\delta}$, for example [9], the fluctuation region 
is wide and then the transition is not sharp, one should use the critical 
fluctuation theory to determine B$_{c2}$(T).

Now we turn to the second transition at T$_{c2}$ on the M(T) curve for the 
overdoped sample. As shown by the open squares in Fig.4, the transition line 
at T$_{c2}$ is extremely positive-curved, being very similar to the 
so-called H$_{c2}$(T) line determined from the resistive measurement 
by Mackenzie et al.[10] in overdoped Tl-2201 samples. As argued in our 
previous paper [5], this transition is not corresponding to the upper 
critical field B$_{c2}$(T), rather to the Josephson coupling [11] or 
proximity effect between the superconducting clusters preformed at 
T$_{c1}$.

By explaining the data  measured on an underdoped and an overdoped 
sample, we derived the following picture: in the overdoped region, 
some superconducting clusters can be formed via electronic phase 
separation. These clusters are surrounded by the good metallic 
regions with rich holes. With lowering temperature, these clusters 
go into superconducting state first at T$_{c1}$ and the bulk 
superconductivity is established probably via Josephson 
coupling or proximity effect between these clusters at 
a lower temperature T$_{c2}$. One may argue that the 
two-step transition for our present overdoped sample is 
induced by some extrinsic causes, for example the possible 
presence of the second chemical phase. This can be however 
ruled out by the observation of very clean ( 00{\it l} ) 
peaks from the x-ray diffraction ( XRD ) pattern even the 
intensity is plotted logarithmically, and the symetric 
non-spliting Laue spots on the present overdoped sample. One 
may furher argue that the presence of a second chemical phase 
is obviously not observed from XRD simpally because of the 
limitations of X-ray diffraction. However, this argument stands 
weakly aginst the same magnitude of diamagnetic moments for these 
two samples as shown in Fig.2. Another possible argument may be 
that there is a nearby first order ( orthorhombic to tetragonal ) 
phase transition at around x = 0.20 as argued in the past by Takagi 
et al.[12], which probably leads to an intrinsic chemical 
inhomogeneity. The two-step transition observed in our overdoped 
sample is certainly not induced by this possible phase transition 
because of following reasons: (1) This two-step behavior has been 
observed both below and above x = 0.20 in the  overdoped region;  
(2) Even this phase transition is truly happened, there is no 
experimental evidence to show that the resultant phase is a mixture 
of two different chemical phases; (3) It has no reason to believe 
that one of the chemical phases ( if exist ),  should have the same 
criticality as the corresponding underdoped sample; (4) The two-step 
behavior has been observed in many different families of overdoped 
samples, for some of them without orthorhombic to tetragonal phase 
transition.       

The picture derived from our measurement inhibits to take the overdoped 
sample as a system with uniformly distributed fermions and thus 
refuses the new theories based on this consideration. As argued 
by Kivelson and Emery [13] that in a system with a local tendency 
to phase separation, one has some kind of  `` Coulomb-frustrated 
phase separation '', i.e., the system is inhomogeneous on an 
intermediate scale. Since these phase separated clusters are 
small, the proximity effect should be operational. In this 
scenario, as the system gets more overdoped, the fraction 
of the superconducting part shrinks and probably the 
typical size of the `` superconducting clusters '' decreases, 
so the T$_{c}$ will be suppressed by the proximity effect, 
but the bulk superconductivity is established via Josephson 
effect or proximity effect at T$_{c2}$. This naturally 
explains the decrease of T$_{c}$ with doping level and the 
second step on the M(T) curve in the overdoped region. In 
regard of inhomoeneity and / or phase separation, our picture 
can get support from substantial recent experiments done on 
overdoped samples, such as overdoped Tl-2201 [10,14,15], 
Bi$_{2}$Sr$_{2-x}$La$_{x}$CuO$_{6+y}$ [5], La$_{2-x}$Sr$_x$CuO$_{4+\delta}$ [16,17], 
and $(Y_{1-x}Ca_x)Ba_2Cu_3O_{7-\delta}$[18,19]. Radcliffe et al.[14] 
measured the electronic specific heat of overdoped Tl-2201 single 
crystals and found that the B$_{c2}$(T) determined from the specific 
heat measurement is much higher than that determined from the 
magnetoresistance measurement. By scaling the electronic specific 
coefficient $\gamma$ values for different doping levels, they 
concluded that the magnetic field forces a portion of the 
superconductor into the normal state, while the remainder in the 
superconducting state is unaffected. This conclusion is fully 
consisten with our picture. Similar anomaly of B$_{c2}$ was 
observed by Blumberg et al. [15] in the measurement of elctronic 
Raman Scattering on overdoped Tl-2201 samples in a high magnetic 
field, although the preliminary explanation to this effect was 
given as the renormalization of quasiparticle spectra in the 
vicinity of vortex lines in the mixed state. Tallon et al. [16] 
reviewed the muon-spin-rotation ( $\mu$SR ) measurement on 
overdoped Tl-2201 and La$_{2-x}$Sr$_x$CuO$_{4+\delta}$ and 
claimed the coexistence of two different regions with different 
superconducting properties arising from the phase separation. 
Similar result was also obtained by Ohsugi et al.[17] in the 
nuclear-magnetic-resonance ( NMR ) measurement in La$_{2-x}$Sr$_x$CuO$_{4+\delta}$ 
and the T$_{c}$ suppression in the overdoped region was attributed to the 
pair-breaking effect associated with a possible structural inhomogeneity. 
Another indirect evidence for the phase separation in the overdoped region 
was from the EXAFS measurement on   $(Y_{1-x}Ca_x)Ba_2Cu_3O_{7-\delta}$ 
samples by Kaldis et al.[18] who concluded that the structure of overdoped  
$(Y_{1-x}Ca_x)Ba_2Cu_3O_{7-\delta}$ samples may be a martensitic form of 
the optimum doped crystals. This may provide a reasonable explanation to 
the two energy gaps found in overdoped  $(Y_{1-x}Ca_x)Ba_2Cu_3O_{7-\delta}$ 
from the time-domain spectroscopic measurement [19]. The general feature of 
inhomogeneity in the overdoped region for various systems as mentioned above 
may hint that the phase separation here is electronically driven, rather than 
due to a chemical or structural segregation, since all these systems have very 
different details of structures. A direct confirmation to our picture would, 
however, come from the scanning-tunneling-microscopic ( STM ) measurement at 
different temperatures under a magnetic field. It can give out the information 
of the spatial electronic density of states provided that the clusters are 
static after phase separation and thus deserves certainly further investigation. 
Our picture may have two folds of impact on theoretical development: Firstly in 
HTS there may be only one pairing mechansim which should get a full reflection 
in the underdoped region, e.g., the pseudogap[20,21] and stripe phase [13], 
etc.. Secondly, any theory for mechanism of HTS should cover an explanation 
to the macroscopic electronic phase separation in the overdoped region. 

\acknowledgments
This work is supported by the Chinese NSFC within the project: 19825111. WHH 
gratefully acknowledges the continuing financial support from the Alexander 
von Humboldt foundation, Germany.


\begin{figure}
\caption{
Temperature dependence of the diamagnetic moments for two typical 
La$_{2-x}$Sr$_x$CuO$_{4}$ single crystals with x = 0.092 ( underdoped ) 
and 0.24 ( overdoped ) measured under an external field of  0.002 T in 
the zero-field-cooling ( ZFC ) process.  In the ZFC process, the sample 
is firstly cooled to a desired temperature at zero field and then an 
external field is applied, the data are collected in the warming up 
process with field. It is clear that the underdoped sample and the 
overdoped sample has equal qualities and roughly equal transition 
temperatures, leading to an effective comparison between these two extreme 
situations.}
\end{figure}

\begin{figure}
\caption{
Temperature dependence of the diamagnetic moments measured for the underdoped 
sample and the overdoped sample at an external field of 1 T in the ZFC and FC 
processes. In the FC process, the sample is firstly cooled to the desired 
temperature under a field and the data are collected in the warming up 
process with field. It is evident that, there is only one transition for 
the underdoped sample but two transitions for the overdoped sample. The 
first transition at T$_{c1}$ for the overdoped sample coincides with the 
solitary transition of the underdoped sample. This transition shifts 
slowly with external field, in sharp contrast to the quickly moved second 
transition at T$_{c2}$ for the overdoped sample.}
\end{figure}

\begin{figure}
\caption{
Temperature dependence of the reversible magnetic moments for the overdoped 
sample at fields of 0.2, 0.4, 1.0, 2.0, 3.5, and 5 T.  A clear common crossing 
point appears at ( T$^*$, M$^*$ ) strongly suggesting an underlying scaling 
behavior. The inset shows the scaling of the data according to eq.(1) by 
taking T$_{c}$ = 25 K, $\alpha$ = 2/3 and dB$_{c2}$ / dT = - 0.7 T / K.} 
\end{figure}

\begin{figure}
\caption{The upper critical field B$_{c2}$ determined for the underdoped sample 
( open circles ) directly from the sharp transition at T$_{c1}$, and for the 
overdoped sample ( solid line ) from the scaling shown in the inset of Fig.3. 
The upper critical field for these two samples are very close to each other 
indicating a same criticality for superconductivity in these two samples albeit 
the overall hole densities are very different. The open squares represent the 
second transition of the overdoped sample and the dotted line is a guide to the 
eyes.}
\end{figure}

\end{document}